# ON BMD TARGET TRACKING:
# DATA ASSOCIATION AND DATA FUSION

Demetrios Serakos

NAVSEA – Dahlgren

Combat Systems Department

Code N14

Dahlgren, VA 22448

(540) 653-1121

ABSTRACT: In this paper we consider multitarget tracking with multiple sensors for BMD. In a previous paper multitarget tracking with a single sensor was considered [8]. A ballistic missile may be in several pieces, presenting multiple targets. Besides the ground based or ship sensor there is also the missile seeker. We consider algorithms for generating and maintaining the tracks needed for BMD. A cue of a BM from a non-organic tracking system may also be received. We consider whether the cue is already in the local track file or is a new track. The cue information can improve the existing local track.

## INTRODUCTION

BMD data association[1] must process measurements from the Ground Based/Ship Radar (RF) and the Missile Seeker (IR).

Data association algorithms for processing RF data may be adjusted to take RF and IR data, see [4] Section 9.6.

BM queuing from a remote source, e.g. another ship, requires remote track to local track fusion[2].

Data association algorithms for processing RF data may be used to associate the remote track with a local track in a multi-track environment. Once associated with a local track, filtering algorithms may be used for the fusion step.

_________________________________________________________________________

1. Data association refers the observation-to-track assignment and track maintenance functions (such as gating) in a multi-target tracking system, see [3] page 6, [4] page 8.

2. Track fusion refers to the two step process of first reconciling two tracks from two different sets of tracks as being of the same object and then combining them into one track, e.g., a remote track fused to a local track, see [3] page 442. The first step is also called track association.

Data association algorithms may be used to cue a missile seeker.

Long track propagation times, which can occur in remote track to local track fusion, may require higher fidelity filtering algorithms.

## PLANT AND OUTPUT EQUATIONS

Plant dynamics follow the Kepler dynamics, see [2]. In ECI coordinates:

$$\frac{d}{dt}\begin{bmatrix} p \\ v \end{bmatrix} = \begin{bmatrix} v \\ -\mu p / |p|^3 \end{bmatrix}$$

Let $x_{ECI} = (p,v)$ '. The above equation is also written as $d\, x_{ECI}/dt = f(x_{ECI})$.

The Jacobian of this equation is: With $p = (x,y,z)$ ' and $v = (v_x, v_y, v_z)$ '.

$$F = \frac{\partial f}{\partial x_{ECI}} = \frac{\partial}{\partial x_{ECI}}\begin{bmatrix} v \\ -\mu p / |p|^3 \end{bmatrix}$$

$$= \begin{bmatrix} 0 & 0 & 0 & 1 & 0 & 0 \\ 0 & 0 & 0 & 0 & 1 & 0 \\ 0 & 0 & 0 & 0 & 0 & 1 \\ -\mu(1/|p|^3 - 3x^2/|p|^5) & 3\mu xy/|p|^5 & 3\mu xz/|p|^5 & 0 & 0 & 0 \\ 3\mu yx/|p|^5 & -\mu(1/|p|^3 - 3y^2/|p|^5) & 3\mu yz/|p|^5 & 0 & 0 & 0 \\ 3\mu zx/|p|^5 & 3\mu zy/|p|^5 & -\mu(1/|p|^3 - 3z^2/|p|^5) & 0 & 0 & 0 \end{bmatrix}$$

For RF measurements, the output equation is range, $r$, azimuth, $\psi$, elevation, $\vartheta$. ENU coordinates $(e,n,u)$ are used. Azimuth is measured from the $n$ axis towards the $e$ axis.

$$z_{RF} = y_{RF} + n_{RF} = \begin{bmatrix} r \\ \psi \\ \theta \end{bmatrix} + n_{RF} = h_{RF}(p_{ENU}) + n_{RF}$$

$p_{ENU} = (e,n,u)$' and $n_{RF}$ is the RF measurement noise.

The Jacobian matrix is:

$$H'_{RF} = \frac{\partial y_{RF}}{\partial x_{ENU}},$$

where $x_{ENU} = (p'_{ENU}, v'_{ENU})$.

$$= \begin{bmatrix} e/r & n/r & u/r & 0 & 0 & 0 \\ -n/(e^2+n^2) & e/(e^2+n^2) & 0 & 0 & 0 & 0 \\ -\left(eu/\sqrt{e^2+n^2}\right)/r^2 & -\left(nu/\sqrt{e^2+n^2}\right)/r^2 & \sqrt{e^2+n^2}/r^2 & 0 & 0 & 0 \end{bmatrix}$$

For IR measurements, the output equation is azimuth, $\psi$ and elevation, $\vartheta$. Missile body coordinates $(x_B, y_B, z_B)$ are used. The standard aircraft convention is used, so that $x_B$ points through the nose of the missile, $y_B$, is in the direction of the right wing and $z_B$ is down. Azimuth is measured by rotation about the $z_B$ axis and elevation is measured by rotation about the $y_B$ axis.

$$z_{IR} = y_{IR} + n_{IR} = \begin{bmatrix} \psi_{IR} \\ \theta_{IR} \end{bmatrix} + n_{IR} = h_{IR}(p_{BODY}) + n_{IR}$$

where $p_{BODY} = (x_B, y_B, z_B)'$ and $n_{IR}$ is the IR measurement noise.

The Jacobian matrix is:

$$H'_{IR} = \frac{\partial y_{IR}}{\partial x_{BODY}}$$

, $x_{BODY} = (p'_{BODY}, v'_{BODY})'$

$$= \begin{bmatrix} -y_B/(x_B^2+y_B^2) & x_B/(x_B^2+y_B^2) & 0 & 0 & 0 & 0 \\ (x_B z_B/\sqrt{x_B^2+y_B^2})/r_B & (y_B z_B/\sqrt{x_B^2+y_B^2})/r_B & -\sqrt{x_B^2+y_B^2}/r_B & 0 & 0 & 0 \end{bmatrix}$$

## TIME-UPDATE USING EKF

With the Extended Kalman Filter (EKF), the first term of Taylor series is used to approximate nonlinear terms.

The mean and covariance of the state vector at the initial time are provided. The mean is propagated $\Delta t$ using the plant dynamics while the covariance is propagated $\Delta t$ using terms from the Taylor series. In these equations, the process noise and measurement noise are assumed uncorrelated.

The mean is time-updated using Euler integration by[1]

$$\begin{bmatrix} \hat{p}_{k+1|k} \\ \hat{v}_{k+1|k} \end{bmatrix} = \begin{bmatrix} \hat{p}_{k|k} \\ \hat{v}_{k|k} \end{bmatrix} + \Delta t \begin{bmatrix} \hat{v}_{k|k} \\ -\mu \hat{p}_{k|k}/|\hat{p}_{k|k}|^3 \end{bmatrix}$$

The covariance is time-updated by

$$\Sigma_{k+1|k} = F_k \Sigma_{k|k} F'_k + Q_k$$

Where $F_k = (I + \Delta t \cdot F)$ and $Q_k$ is the process noise intensity matrix for the plant.

## TIME UPDATE USING UNSCENTED KF

With the Unscented Kalman Filter (UKF) see [6] and [7], a small number of points selected about the perimeter of the error basket (a region about the mean defined by one-sigma covariance) are propagated $\Delta t$ using the plant dynamics. These propagated points are used to construct the propagated mean and covariance.

Given a 6 by 1 mean vector $(p,v)'$ and a 6 by 6 covariance matrix $\Sigma$ with Cholesky factorization

$\Sigma = LL'$ where $L = [l_1, l_2, l_3, l_4, l_5, l_6]$ (6 by 6) is lower triangular define the points

$$X_{i,k|k} = \hat{x}_{k|k,ECI} \pm W_i \cdot l_{i,k}$$

where $W_i$ are weighting factors so that after the reconstruction process the propagated covariance is the right size. Set $X_{0,k|k} = \hat{x}_{k|k,ECI}$.

Propagate in time the $L$ points using the plant dynamics.

$$X_{i,k+1|k} = f(X_{i,k|k}), i = 0,1,\cdots,12$$

Reconstruct the mean at the final time:

$$\hat{x}_{k+1|k,ECI} = \sum_{i=0}^{12} W_i X_{i,k+1|k}$$

Reconstruct the covariance at the final time:

$$\Sigma_{k+1|k} = \sum_{i=0}^{12} W_i \cdot (X_{i,k+1|k} - \hat{x}_{k+1|k,ECI})(X_{i,k+1|k} - \hat{x}_{k+1|k,ECI})^T$$

## DATA ASSOCIATION WITH RF AND IR MEASUREMENTS

During a BMD engagement, in mid-course the BM is tracked with the ship's radar. In the terminal phase, the BM is tracked by the missile IR seeker as well.

In this paper the maximum likelihood method is presented. Although, other data association methods such as Multiple Hypothesis Tracking (MHT) method and the Joint Probability Data Association (JPDA) method of Bar-Shalom, may be used, see [8].

These three methods may be adapted to incorporate IR data. When RF and IR data are both available, the output equation is generated by augmenting the RF and IR equations:

$$z_{AUG} = \begin{bmatrix} z_{RF} \\ z_{IR} \end{bmatrix} = \begin{bmatrix} h_{RF}(x_{ENU}) + n_{RF} \\ h_{IR}(x_{BODY}) + n_{IR} \end{bmatrix}$$

We assume that the RF and IR sensors are synchronized sensors - or else $z_{AUG}$ would not be available. (If these sensors are not synchronized, in the simplest case where the measurements are chronologically available, the processing would be done as the data became available.)

## DATA ASSOCIATION WITH RF AND IR MEASUREMENTS

Sequential processing is a term used in Kalman filtering for a technique where the output is partitioned and the measurement updates for the partitions are sequentially processed. The RF and IR measurement noise must be independent. The measurements $z_{RF}$ and $z_{IR}$ each update the state.

Maximum likelihood association review, see [8] : Given $m$ measurements and $n$ tracks. An $m$ by ($m+n$) association matrix is computed. The $i, j$ component of this matrix is the value of fitting the $i$th measurement to the $j$th track if $i \leq n$ or starting a new tentative track for the measurement when $i > n$. The main factor in determining this value is the residual between the $i$th measurement and the $j$th track. An assignment algorithm, such as Bertsekas' auction algorithm, is used to maximize the assignments over all measurements. Each measurement is assigned uniquely to a track. Track scores are computed for each track which depend on the association value, track length, number of updates, etc. If a measurement is associated to a track, the score goes up, the score goes down when no new measurement is associated to it. Tracks are deleted if their scores drop too low, tentative tracks become confirmed tracks when their scores become high enough relative to a threshold value.

The sequential processing approach can be adapted to the multi sensor data association problem. Consider the following procedure: Assume a centralized architecture; that is, the RF and IR data are sent to a centralized location where the tracks are maintained and stored. Given $m_{RF}$ RF measurements, $m_{IR}$ IR measurements and $n$ tracks, form an $m_{RF}$ by $n$ association matrix and an $m_{IR}$ by $n$ association matrix. The association algorithm is performed on each of these matrices. A particular track may have at most one RF and one IR measurement assigned to it, i.e. no measurement can be used with more than one track. Measurement update the tracks that have an RF measurement assigned to them. Measurement update the tracks that have an IR measurement assigned to them. RF and IR track scores are maintained. Total track score is the sum of these.

The above association method may be used to associate the RV track from the RF senor to the IR measurement from the IR seeker. Assume the RF sensor has completed the discrimination process and identified the RV, thus providing a pointing vector to the IR seeker. In this case; The IR seeker makes observations of objects in its field of view. The association matrix is then computed with the IR observations. The IR data associated with RV identified track is used as a pointing target for the IR seeker. If no IR data is associated with the RV identified track, a missing return is declared and the RV track is coasted.

## MEASUREMENT UPDATE, EKF

We pick-up at the point where each track has at most one RF measurement and one IR measurement assigned to it.

By choice, block sequential process the RF data (if any) followed by the IR data (if any).

Process the RF data for the EKF: The RF Kalman gain is:

$$\Lambda_{k,RF} = \Sigma_{k|k-1} T_{ENU}^{ECI} H_{k,RF} \cdot (H'_{k,RF} T'^{ECI}_{ENU} \Sigma_{k|k-1} T_{ENU}^{ECI} H_{k,RF} + R_{RF})^{-1}$$

($T_{ENU}^{ECI}$ is the 6 by 6 ENU to ECI transformation matrix with $E[n_{RF} n'_{RF}] = R_{RF}$)

Measurement update the covariance with the RF measurements:

$$\Sigma_{k|k,RF} = \Sigma_{k|k-1} - \Lambda_{k,RF} H'_{k,RF} T'^{ECI}_{ENU} \Sigma_{k|k-1}$$

Measurement update the state with the RF measurements:

$$\hat{x}_{k|k,RF,ECI} = \hat{x}_{k|k-1,ECI} + \Lambda_{k,RF}(z_{k,RF} - T'^{ECI}_{ENU} h_{RF}(\hat{x}_{k|k-1,ECI}))$$

Process the IR data for the EKF: The IR Kalman gain is (with $E[n_{IR} n'_{IR}] = R_{IR}$):

$$\Lambda_{k,IR} = \Sigma_{k|k,RF} T_{BODY}^{ECI} H_{k,BODY} \cdot (H'_{k,BODY} T'^{ECI}_{BODY} \Sigma_{k|k,RF} T_{BODY}^{ECI} H_{k,BODY} + R_{IR})^{-1}$$

Measurement update the state with the IR measurements.

$$\hat{x}_{k|k,ECI} = \hat{x}_{k|k,RF,IR,ECI}$$

$$= \hat{x}_{k|k,RF,ECI} + \Lambda_{k,IR}(z_{k,IR} - T'^{ECI}_{BODY} h_{IR}(\hat{x}_{k|k,RF,ECI}))$$

Measurement update the covariance with the IR measurements.

$$\Sigma_{k|k} = \Sigma_{k|k,RF,IR} = \Sigma_{k|k,RF} - \Lambda_{k,IR} H'_{k,IR} T'^{ECI}_{BODY} \Sigma_{k|k,RF}$$

## MEASUREMENT UPDATE, UKF

Process the RF data for the UKF: Compute the predicted observation: (Note this is different from $h_{RF}(\hat{x}_{k+1|k,ECI})$.)

$$\zeta_{k+1|k,RF} = \sum_{i=1}^{12} W_i h_{RF}(X_{i,k+1|k})$$

Compute the innovation covariance matrix: (Note that the sum goes from $i = 0$ even though $\zeta_{k+1|k,RF}$ was computed by summing from $i = 1$.)

$$\Sigma_{zz,k+1|k} = \sum_{i=0}^{12} W_i (h_{RF}(X_{i,k+1|k}) - \zeta_{k+1|k,RF})(h_{RF}(X_{i,k+1|k}) - \zeta_{k+1|k,RF})^T$$

Compute the state - measurement cross correlation matrix:

$$\Sigma_{xz,k+1|k} = \sum_{i=0}^{12} W_i (X_{i,k+1|k} - \hat{x}_{k+1|k,ECI})(h_{RF}(X_{i,k+1|k}) - \zeta_{k+1|k,RF})^T$$

Measurement update similar to EKF. Compute Kalman gain:

$$\Lambda_{k,RF} = \Sigma_{xz,k|k-1} \Sigma_{zz,k|k-1}^{-1}$$

Measurement update state with the RF measurements:

$$\hat{x}_{k|k,ECI} = \hat{x}_{k|k-1,ECI} + \Lambda_{k,RF}(z_{k,RF} - \zeta_{k,RF})$$

Measurement update covariance with the RF measurements:

$$\Sigma_{k|k,RF} = \Sigma_{k|k-1} - \Lambda_{k,RF} \Sigma_{zz,k|k-1} \Lambda'_{k,RF}$$

Process the IR data for the UKF: Similar the EKF process.

### REMOTE TRACK TO LOCAL TRACK FUSION

Remote BM tracks may be associated and fused with local BM tracks in a BM engagement. This is a two step process. Determine to which local track the remote track is associated – the association step. Combine the means and covariances of these tracks – the fusion step. The association and fusion techniques mentioned earlier in this paper may be adapted to this problem.

Two methods for track fusion are presented. The first fuses the tracks by taking a linear combination of the tracks. The second method views the remote track as a measurement and uses the EKF (or UKF).

Sun and Deng, [9], present a method for fusing unbiased estimates. Let $\hat{x}_R, \hat{x}_L; \Sigma_R, \Sigma_L$ be local and remote unbiased estimators of a random vector $x$, along with their corresponding covariances. An unbiased estimator is computed by taking a linear combination of the given estimators.

$$\hat{x}_F = A_R \hat{x}_R + A_L \hat{x}_L$$

A performance function is defined: (with $\tilde{x}_F = x_{ECI} - \hat{x}_F$ )

$$J = \text{tr}(E[\tilde{x}_F \tilde{x}_F^T])$$

Lagrange multipliers are used to minimize the performance function subject to the constraint $A_R + A_L = I_6$. This results in:

$$\begin{bmatrix} \overline{A}_R \\ \overline{A}_L \end{bmatrix} = \begin{bmatrix} \Sigma_R & \Sigma_{RL} \\ \Sigma_{LR} & \Sigma_L \end{bmatrix}^{-1} \begin{bmatrix} I_6 \\ I_6 \end{bmatrix} \left( \begin{bmatrix} I_6 & I_6 \end{bmatrix} \begin{bmatrix} \Sigma_R & \Sigma_{RL} \\ \Sigma_{LR} & \Sigma_L \end{bmatrix}^{-1} \begin{bmatrix} I_6 \\ I_6 \end{bmatrix} \right)^{-1}$$

If $x_R$ is independent of $x_L$, that is $\Sigma_{RL} = 0$, the above equations reduce to

$$\hat{x}_F = \left( \Sigma_R^{-1} \hat{x}_R + \Sigma_L^{-1} \hat{x}_L \right) \left( \Sigma_R^{-1} + \Sigma_L^{-1} \right)^{-1}$$

From [3] page 447, we get that $x_R$ and $x_L$ are correlated because of common process noise. Note that $\Sigma_{RL}$ will be negligible if Kepler Dynamics are used which have low process noise. The covariance of the fused track is

$$\Sigma_F = \left( \begin{bmatrix} I_6 & I_6 \end{bmatrix} \begin{bmatrix} \Sigma_R & \Sigma_{RL} \\ \Sigma_{LR} & \Sigma_L \end{bmatrix}^{-1} \begin{bmatrix} I_6 \\ I_6 \end{bmatrix} \right)^{-1}$$

A second approach to fusing the remote and the local tracks is to view the remote track as a measurement. This is similar to [3] Section 8.3.3. The plant equation is as described above. The output equation is:

$$z_F = \hat{x}_R = x + n_R$$

Consider the statistics for this problem. The measurement noise covariance is set to the covariance of the remote track: $R_R = E[n_R n'_R] = \Sigma_R$. The covariance of the process noise is as in the previous case: $E[n_L n'_L] = Q_L$. Since the measurement is an estimate of the state from a remote sensor, we need to assume that the process noise and measurement noise are correlated. Hence, set $E[n_R n'_L] = S_{RL}$. Use the EKF measurement update equation to fuse the tracks:

$$\hat{x}_{k|k,F} = \hat{x}_{k|k-1,L} + \Sigma_{k|k-1,L} (\Sigma_{k|k-1,L} + \Sigma_{k|k-1,R})^{-1} (z_F - \hat{x}_{k|k-1,L})$$

The covariance of the fused remote and local tracks is computed with the EKF covariance update equation:

$$\Sigma_{k|k,F} = \Sigma_{k|k-1,L} - \Sigma_{k|k-1,L} (\Sigma_{k|k-1,L} + \Sigma_{k|k-1,R})^{-1} \Sigma_{k|k-1,L}$$

The time update to be used after the fusion step (because of $S_{RL}$) for the state and covariance are: ($x_F$ and $\Sigma_F$ denotes the fused estimate and covariance.)

$$\hat{x}_{k+1|k,ECI} = \hat{x}_{k+1|k,F} = f(\hat{x}_{k|k,ECI}) - S_{RL} R_R^{-1} \hat{x}_{k|k,ECI} + S_{RL} R_R^{-1} z_F$$

$$\Sigma_{k+1|k} = \Sigma_{k+1|k,F} = (F - S_{RL} R_R^{-1}) \Sigma_{k|k} (F - S_{RL} R_R^{-1})' + (Q_L - S_{RL} R_R^{-1} S'_{RL})$$

After the fusion step, go back to the regular equations. Two methods can be used to compute $S_{RL}$ as given by: Bar-Shalom and Li, [3] page 448, provide a method which may be of use in calculating $S_{RL}$. Chen et. al., [5], give a method which may be useful in selecting a worst case $S_{RL}$ when $S_R$ and $S_L$ are known. The error basket of the fused tracks should contain the intersection of the remote and local error baskets.

## SUMMARY AND CONCLUSIONS

This paper discusses the RF/IR data association. By augmenting the RF output equation with the IR measurements, the previously discussed maximum likelihood method may accommodate RF and IR data. Both the Extended Kalman Filter and the Unscented Kalman Filter can be used for the filtering process.

If the lethal object has been identified, the given algorithm may be used to point the missile seeker towards the RV.

Two methods for remote track to local track fusion were presented. The first method took a linear combination of the remote and local tracks. The second method used the remote track as a measurement in standard filtering algorithms. In a multi-track environment the remote track may be associated with the correct local track by treating the remote track as a measurement.